\documentstyle[epsfig,aps,prl,multicol]{revtex}
\begin{document}
\draft
\title{Bandgap renormalization of  modulation doped quantum wires}
\author{S. Sedlmaier, M. Stopa, G. Schedelbeck, W. Wegscheider$^*$ and G. Abstreiter} 
\address{
Walter Schottky Institut \\ Am Coulombwall \\ D-85748 Garching, Germany \\
phone 49-89-289-12198, Fax: 49-89-289-12737 \\ e-mail: stopa@wsi.tum.de}
\date{\today}
\maketitle

\begin{multicols}{2}
[
\begin{abstract}
We measure the photoluminescence (PL) spectra for an array
of modulation doped, T-shaped quantum wires as a function of the 1d
density $n_e$ which is modulated with a surface gate. 
We present 
self-consistent electronic structure calculations for this device
which show a bandgap renormalization which, when corrected for 
excitonic energy and its screening, are largely insensitive to $n_e$
and which are in quantitatively excellent agreement with the data.
The calculations show that electron and hole remain bound up to
$\sim 3 \times 10^6 \; cm^{-1}$ and that therefore the stability of the
exciton far exceeds the conservative Mott criterion.
\end{abstract}
]

Exchange and correlation in an electron gas 
formed in a semiconductor act to counter the direct Coulomb 
interaction by reducing the inter-particle overlap. For
two component systems, such as the electron-hole plasma
created in optical experiments, this effect tends to
produce a bandgap renormalization (BGR) with increasing 
density $n_e$ and/or $n_h$, which reduces
the energy of photons emitted upon recombination from
the band edges \cite{reviews}. Exciton formation further reduces
the bandgap but exciton binding is weakened
by mobile charges and so the trend with density opposes that
of exchange-correlation induced BGR.
Investigations of the bandgap, which
has a pivotal dependence on the dimensionality of the system, 
is of interest both for its significance to optical technology and
for the illumination it provides for the many-body problem.
Consequently there are numerous experimental and theoretical
studies of BGR which have focused on systems of successively lower
dimension over the past several years \cite{Sch89,Das89}.

For one dimensional systems, or quantum wires (QWRs), a number of
recent experimental and theoretical accounts have begun to
clarify the often competing effects which result in
density dependent changes to the observed photoluminescence 
energy \cite{Weg93,Amb97,Cin93,Ben92,Ben91,Hwa98}.
In general, BGR depends on the densities, $n_e$ and $n_h$, 
of both components of the electron-hole plasma. Typically, however,
research has concentrated on intrinsic samples wherein $n_e=n_h=n$.
One difficulty with this approach has been that in order to
vary $n$, an increase in photoexcitation has been required,
or else the time development as the excitation subsides has been
observed, and the resulting spectra are complicated with highly
non-equilibrium effects such as phase space filling.
In this paper we present an experimental study of the
evolution of the photoluminescence energy in a doped, T-shaped
QWR sample whose conduction band electron density $n_e$ can be modulated 
with the voltage applied to a surface gate. This structure
(see Fig. 1), fabricated via the cleaved edge overgrowth
techniques \cite{Pfe90} has appreciable advantages. First it provides
wires of high precision, with structural variations restricted
to the monolayer regime. Second, it permits the comparison of
wire and quantum well photoluminescence in a single sample.

To complement the measurements, we
perform self-consistent electronic structure calculations within
density functional theory (DFT) for this structure, using the local density 
approximation (LDA)
for exchange and correlation (XC), $V_{xc}$.
The theoretical bandgap renormalization, which is usually
calculated with many-body techniques, 
is equivalent to the {\it difference} between the
LDA calculated bandgap and that calculated within a pure Hartree 
approximation, which omits the $V_{xc}$ term. We have further 
calculated the effect of exciton formation and its screening
on the bandgap, using a simplified model potential with
parameters derived from the (translationally invariant) DFT
calculation.

\begin{center}
\begin{figure}
\begin{minipage}{8cm}
\epsfig{file=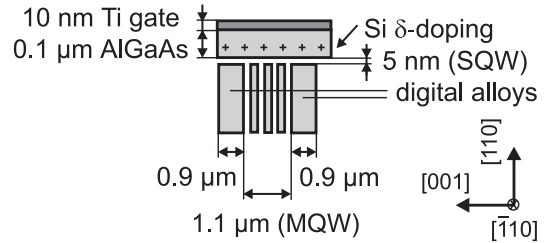,width=7cm}
\vspace*{2mm}
\caption{Sample schematic showing the multiple, parallel T-shaped QWRs
formed at the intersection of the edges of the
multiple quantum wells (MQWs)
with the single, modulation doped quantum well (SQW) 
in the overgrowth layer (not true scale).}
\end{minipage}
\end{figure}
\end{center}

Our principle result is that, as with experiments on two-component
plasma in V-groove wires \cite{Amb97}, the photoluminescence peak 
position is largely insensitive to density.  The calculated screening 
of the exciton reduces the  binding energy with a functional form that 
neatly cancels most of the XC induced BGR, predicting a recombination
energy in excellent agreement with experiment. Additionally, the
appearance of sharp structure in the PL data, indicating recombination
from excitons localized at monolayer potential fluctuations,
which gradually vanishes with increasing $n_e$, supports this
BGR+exciton screening model. The calculation suggests that
the exciton remains bound for very high density, also in
agreement with Ref. \cite{Amb97}, however the
sharp structure disappears at much lower density, 
$n_e \sim 1 \times 10^6 \; cm^{-2}$,
indicating delocalization of the exciton.

The cleaved edge overgrowth (CEO) technique employed for our QWR
structure has been described in detail elsewhere \cite{Pfe90}. Our 
structure consists of 22 periods of (001)-oriented 
$GaAs$ ($5 \; nm$) / $Al_{0.32}Ga_{0.68}As$ ($44 \;nm$) 
quantum wells (multiple quantum wells, MQWs), grown between two digital 
alloys with 90 periods of $GaAs$ ($2 \; nm$) / $Al_{0.32}Ga_{0.68}As$ 
($8 \; nm$) each. These digital alloys permit us to observe the PL from 
the overgrowth single quantum well (SQW), which is defined by growing 
along the [110]-crystal axis $5 \; nm$ $GaAs$, a $30 \; nm$ 
$Al_{0.35}Ga_{0.65}As$ spacer, a silicon $\delta$-doping 
(n-modulation doping), and $70 \; nm$ $Al_{0.35}Ga_{0.65}As$. 
After both growth steps, 10 nm thick cap layers are added, which
are not included in Fig. 1.

\begin{center}
\begin{figure}
\begin{minipage}{8cm}
\epsfig{file=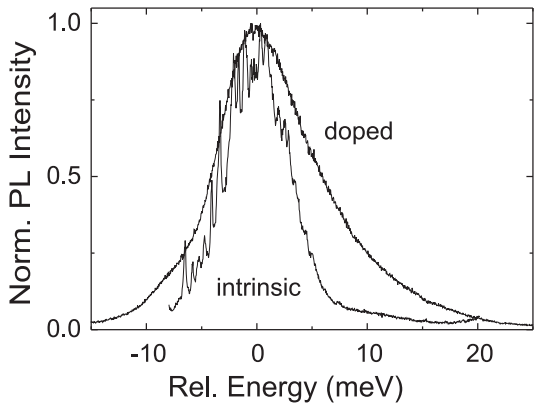,width=7cm}
\vspace*{2mm}
\caption{Comparison between the normalized PL lineshapes of intrinsic 
and n-modulation doped T-shaped QWRs ($n \sim 1 \times 10^6 \; cm^{-1}$). 
The energy axis is set with
respect to the peak position of the respective PL line.
Excitation and detection polarizations are chosen parallel to the QWRs, 
consistent with Fig. 3. In order to detect only the QWRs PL and to 
avoid an overlap with the SQW PL, as in Fig. 3, excitation is performed
on the ($110$)-surface. The excitation power amounts 
to 1 $\mu W$, the excitation energy to 1656 meV.}
\end{minipage}
\end{figure}
\end{center}
T-shaped QWRs form with atomic
precision at each $5 \times 5 \; nm^2$ wide intersection of the 
SQW with one of the
multiple quantum wells \cite{Glu97,Gon92}. 

In order to continuously vary the electron density in the QWRs and the SQW, 
we evaporate a $10 \; nm$ thick, semi-transparent titanium gate on the 
surface of the overgrowth layer of
a second set of samples.
When the gate is grounded, the electron density in the SQW and in the
QWRs are close to that of the un-gated samples.

To maximize spatial
resolution of the photoluminescence (PL) and photoluminescence excitation 
(PLE) spectroscopy, we focus the
excitation beam of a tunable cw dye laser, pumped by an $Ar$-ion laser,
with a microscope objective onto the sample, which is attached inside a
cryostat to a copper block at the nominal temperature $5 \; K$.  On the 
sample, the diameter of
the almost diffraction-limited laser spot amounts to about $800 \; nm$
full-width at half-maximum.  A confocal imaging system guarantees that
only PL limited to the laser spot region is detected.

For both un-gated and gated samples, PLE spectra reveal that, due to an 
electron transfer from the doping layer
into the SQW, an electron system is generated both in the SQW, between
digital alloy and overgrowth spacer, and in the QWRs.

Exciting an un-gated sample on the ($\bar{1}10$)-surface, we are able to
identify the QWRs PL because it is localized exactly and exclusively
at the intersecting region of single and multiple quantum wells and is
emitted at lower energy than the PL of the SQW and the MQWs \cite{Som96}.  
Of course,
individual QWRs cannot be resolved, since they are spaced by $44 \; nm$ only,
with respect to a spatial resolution of our
instrument of about $800 \; nm$.  In order to
interpret the PL lineshape of the n-modulation doped T-shaped QWRs, we
compare it with the lineshape of {\it intrinsic} T-shaped QWRs, as shown in
figure 2 (the curves are aligned horizontally so that the peaks coincide).
Interface roughness,
particularly in the (110)-oriented single quantum well \cite{Gro94}, results
in an inhomogeneously broadened, on average symmetric PL line of the
intrinsic QWRs.  The spectrally sharp peaks on
the PL line are attributed to excitons localized at monolayer potential fluctuations 
\cite{Has97}.  In the presence of free carriers, however, the excitonic
electron-hole interaction is screened \cite{Ros96} and the sharp peaks on the
PL line disappear in the case of n-modulation doped QWRs.  
Furthermore, the asymmetric PL lineshape indicates the
formation of an 1D electron plasma in our modulation doped QWRs and, as
the dominating recombination mechanism, band-to-band transitions between
electrons of the Fermi sea and photogenerated holes. The 
Maxwell-Boltzmann distribution of the photogenerated holes and the
joint one-dimensional density of states $ \sim 1/ \sqrt{E}$ \cite{Cin93} 
result in a
decreasing recombination rate with increasing transition energy, if we
assume a constant transition matrix element for k-conserving band-to-
band recombinations \cite{Cin93}.  In k-space, these transitions occur from
the $\Gamma$-point of the first Brillouin zone up to the Fermi wave vector,
if we presume low temperatures. The decrease of the recombination rate with
increasing transition energy is interpreted as the origin of the wide
high energy tail and therefore the asymmetry of the PL line for the
modulation doped QWRs.  On the other hand, observing band-to-band
transitions means that the electron density in the QWRs, whose charge
density is estimated at $1 \times 10^6 \; cm^{-1}$ \cite{Sto99}, 
exceeds the Mott density
\cite{Ros96}.  According to simulation results \cite{Sto99}, 
only electrons in the
first QWRs subband have maximum probability density at the
T-intersections.  Electrons in the second subband are localized
principally between pairs of T-intersections and have inappreciable overlap
with the hole subbands. Therefore recombinations from higher subbands
can be excluded as the origin of
the asymmetric PL lineshape.

\begin{center}
\begin{figure}
\begin{minipage}{8cm}
\epsfig{file=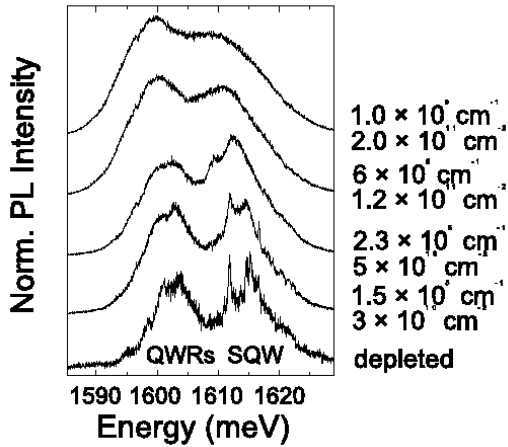,width=7cm}
\vspace*{2mm}
\caption{Normalized and offset PL spectra of n-modulation 
doped QWRs and the SQW. We have
increased the excitation power 
to 11 $\mu W$  because of the lower PL intensity in this geometry. 
The excitation energy is 1645 meV, the detected 
polarization lies parallel to the QWRs.}
\end{minipage}
\end{figure}
\end{center}

For the gated sample, excitation is performed from the (001)-sample surface. This
permits us to observe simultaneously and compare the PL 
(figure 3) of the QWRs (peak on low energy side) and the SQW (high side) between
digital alloy and overgrowth spacer. The MQWs PL occurs at higher energy than
the exhibited energy window.  Applying a negative gate voltage relative to
the electron system, the charge density in the QWRs and in the SQW is reduced. In
figure 3, we have converted applied gate voltage into electron density
per unit length for the QWRs and per unit area for the SQW. 
The bottom spectra of figure 3
displays the response for complete depletion
as confirmed by a series of PLE
measurements.  If the depletion voltage is further increased, 
neither the PL peak position, 
nor qualitatively the PL lineshape change, which is consistent with a total 
depletion of the electron systems for the bottom spectra.
With decreasing electron density, the PL lines
of both the 
QWRs and the SQW narrow slightly, which is consistent with a reduction of the 
Fermi wave vector for both the QWRs and the SQW.  Note that the estimated
electron density of $2 \times 10^{11} cm^{-2}$ 
in the SQW exceeds the 2D Mott-density
\cite{Bau91}.  At low densities, moreover, sharp peaks appear on the PL
lines, which we attribute to excitonic, spatially localized
recombination.  For the SQW (figure 4), one 
\begin{center}
\begin{figure}
\begin{minipage}{8cm}
\epsfig{file=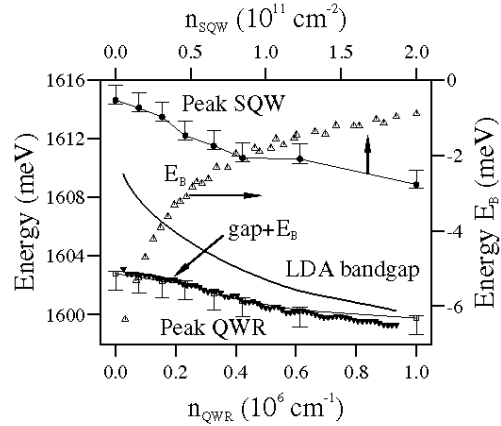,width=7cm}
\vspace*{2mm}
\caption{Electron density dependence of the energetic peak positions 
of the QWRs (squares) and the SQW (dots, upper density scale), 
obtained by a simple 
lineshape fit of the spectra in figure 3. 
Error bars indicate the uncertainty in determining the 
peak positions. The calculated LDA bandgap (heavy line)
and the exciton binding energy $E_B$ (hollow triangles, right scale)
combine to produce the corrected gap (solid inverted triangles)
which fit the data remarkably well. Note an overall offset is arbitrary due to
the choice of the bulk bandgap.}
\end{minipage}
\end{figure}
\end{center}
notices a characteristic redshift 
of the energetic peak position of about $5 - 6 \; meV$, obtained by a simple
lineshape fit, as the 2d electron density, $N_e$, is increased from zero 
to about $2 \times 10^{11} cm^{-2}$.  Correcting for the energy shift
due to the quantum confined Stark effect \cite{Bas88b}, determined by
self-consistently solving Schr\"{o}dinger´s and Poisson´s equation,
the residual shift, due to 2D BGR,
amounts to about $5 \pm 1 \; meV$, in good agreement with earlier results for
a n-modulation doped quantum well \cite{Del87}.  The indicated tolerance
takes into account the uncertainty in determining the real PL peak
position, as two PL lines overlap in figure 3.  

The principle result of figure 4, however, is the weak variation of
the peak position for the QWRs as $n_e$ varies. The overall
shift of only about $3 \; meV$, when the electron density is 
increased from zero to about $1 \times 10^6 cm^{-1}$, is 
similar to results found for wires with a two component
plasma in high excitation \cite{Amb97}.

The observation is in excellent agreement with the variation
of the bandgap determined by the LDA calculation when the
excitonic screening correction is included. The
details of our calculation, which are based on a total
free energy functional for the interacting wire-gate
system \cite{LP1} will be discussed in a separate 
publication \cite{Sto99}. In figure 4 we plot the variation
of the translationally invariant band edge, the calculated
exciton binding energy and the combination of the
two as a function of $n_e$. Clearly, without the screening
of the exciton, the band edge variation disagrees
markedly with measurement. The variation of the exciton
binding, however, is functionally nearly the inverse of 
the band edge variation, with variation at
low $n_e$ strongest in both cases. The result is a close 
cancellation and a trend with $n_e$ that recapitulates 
the data.

A similar cancellation of exciton binding energy and 
BGR has been derived recently by Das Sarma and Wang \cite{Das99}
using the Bethe-Salpeter equation, for the case of a two-component,
neutral plasma (i.e. for $n_e=n_h$). However one striking
contrast between our results and those of 
Ref. \cite{Das99} is that, up to our highest 
density $n_e = 3 \times 10^6 \; cm^{-1}$, we find that the electron
and hole remain bound (cf. Fig. 4), whereas those authors find
a merging of the exciton with the continuum, a so-called
``Mott transition,'' in the range of $0.3 \times 10^6 \; cm^{-1}$. 
The robustness of the exciton revealed in our calculations
emerges from the requirement of
orthogonality between the free, screening electrons and those
bound to the hole; a constraint which is not maintained in
the many-body calculation. Therefore, at least in the case of a 
one component plasma, we find that the stability of the exciton 
exceeds that predicted by the conservative Mott criterion.

In addition our calculation
employs the full non-linear screening, whereas the many-body
calculation \cite{Das99} assumes linear screening and hence is not 
valid in the low density limit. Regarding this point, it is in
the low density regime where BGR and the excitonic binding energy
change most rapidly with density. The result that there remains
a strong tendency for the two effects to cancel is therefore
suggestive of a fundamental connection between the two
processes. The exchange portion of the energy, which 
dominates $V_{xc}$ at low density, varies as $-\rho^{1/3}$.
Therefore a $+\rho^{1/3}$ dependence for the screened
exciton interaction is suggested, although we do not have
a fundamental argument for this.

In conclusion, we have presented photoluminescence measurements
of a modulation doped and surface gated T-shaped quantum wire 
which exhibit a weak dependence of the peak position on
the density of conduction band electrons in the wire. We
have also reported on density functional calculations for the
structure which show a bandgap renormalization of $\sim \, -10 \; meV$
over the range of measured densities. A calculation
of the excitonic binding energy and its screening shows a
complementary trend to the BGR such that the combined results
are largely insensitive to $n_e$ and agree well with the
observed line peak. Finally, we find that while the exciton
binding weakens with density, it nonetheless remains bound up to 
$n_e = 3 \times 10^6 \; cm^{-1}$, suggesting an excitonic stability
well in excess of the Mott criterion.

$*$ permanent address: Universit\"{a}t Regensburg, 93040 Regensburg,
Germany

\end{multicols}

\end{document}